\input{aipcheck}

\documentclass[
    ,final            
  ]
  {aipproc}

\layoutstyle{6x9}

     \newcommand{\la}{\,\rlap{\raise 0.5ex\hbox{$<$}}{\lower 1.0ex\hbox{$\sim$}}\,}
     \newcommand{\ga}{\,\rlap{\raise 0.5ex\hbox{$>$}}{\lower 1.0ex\hbox{$\sim$}}\,}
\newcommand\fd{\hbox{$.\!\!^{\reset@font\romn d}$}}
\newcommand\fh{\hbox{$.\!\!^{\reset@font\romn h}$}}
\newcommand\fm{\hbox{$.\!\!^{\reset@font\romn m}$}}
\newcommand\fs{\hbox{$.\!\!^{\reset@font\romn s}$}}

\newcommand\fp{\hbox{$.\!\!^{\reset@font\reset@font\scriptscriptstyle\romn p}$}}

\def \sw {{\em Swift}}
\def \apj {ApJ}
\def \apjl {ApJL}

\def \mnras {MNRAS}

\begin{document}

\title{Investigating Supergiant Fast X-ray Transients with LOFT}


\author{P.~Romano}{
 address={INAF - IASF Palermo,
       Via U.\ La Malfa 153, I-90146 Palermo, Italy}
}

\author{E.~Bozzo}{
  address={ISDC, University of Geneva, Chemin d'\'Ecogia 16, 1290 Versoix, Switzerland}
}

\author{P.~Esposito}{
  address={INAF -IASF Milano, 
        Via E.\ Bassini 15,   I-20133 Milano,  Italy}
}

\author{C.~Ferrigno}{
  address={ISDC, University of Geneva, Chemin d'\'Ecogia 16, 1290 Versoix, Switzerland}
}
 
\author{V.~Mangano}{
 address={INAF - IASF Palermo,
       Via U.\ La Malfa 153, I-90146 Palermo, Italy}
}

\begin{abstract}
Supergiant Fast X-ray Transients (SFXT) are a class of High-Mass X-ray 
Binaries whose optical counterparts are O or B supergiant stars, 
and whose X-ray outbursts are about 4 orders of magnitude brighter 
than the quiescent state. 
LOFT, the Large Observatory For X-ray Timing, with its coded mask 
Wide Field Monitor (WFM) and its 10 m$^2$ class collimated X-ray 
Large Area Detector (LAD), will be able to dramatically deepen the
knowledge of this class of sources. It will provide simultaneous
high S/N broad-band and time-resolved spectroscopy in several
intensity states, and long term monitoring that will yield new 
determinations of orbital periods, as well as spin periods. 
We show the results of an extensive set of simulations performed 
using previous observational results on these sources obtained with \sw\  and {\it XMM-Newton}.
The WFM will detect all SFXT flares within its field of view 
down to a 15--20 mCrab in 5\,ks. 
Our simulations describe the outbursts at several intensities 
($F_{\rm (2-10\,keV)}=5.9\times10^{-9}$ to $5.5\times10^{-10}$ erg cm$^{-2}$ 
s$^{-1}$), the intermediate and most common state ($10^{-11}$ erg 
cm$^{-2}$ s$^{-1}$), and the low state ($1.2\times10^{-12}$ to 
$5\times10^{-13}$ erg cm$^{-2}$ s$^{-1}$). We also considered large 
variations of $N_{\rm H}$ and the presence of emission lines, as observed 
by {\it Swift} and {\it XMM--Newton}.
\end{abstract}

\maketitle


             \section{Perspectives for Supergiant fast X-ray transients}

Supergiant fast X-ray transients (SFXTs) are High-Mass X-ray 
Binaries (HMXB) transients that show flares peaking at 
$10^{36}$--$10^{37}$ erg s$^{-1}$ (2--10 keV observed flux of $10^{-9}$ erg cm$^{-2}$ s$^{-1}$) 
lasting a few hours. Their X-ray spectra during a flare are similar to those of accreting neutron stars, 
and can be described as a flat power-law below $\sim10$ keV sometimes displaying a cutoff at 15--30 keV.  
Their large X--ray dynamic range (3--5 orders of magnitude) and their association with 
OB supergiant companions make them a peculiar class of HMXBs that now include  about 
10 SFXTs and 10 candidates (showing the same X--ray properties but still lacking an 
optical spectroscopic classification). 

LOFT, the Large Observatory For X--ray Timing \cite{FerociLOFT_mn}, is a newly proposed 
space mission selected by ESA in February 2011 as one of the four M3 mission candidates that will compete 
for a launch opportunity at the start of the 2020s.
The LOFT Large Area Detector \cite[LAD,][]{Zane2012LAD_mn} has an effective area a
 factor of $\sim20$ larger than RXTE/PCA 
(the largest area X--ray instrument ever flown) and much improved energy resolution 
(better than 260\,eV). It is specifically designed to exploit the diagnostics of very rapid 
X-ray flux and spectral variability that directly probe the motion of matter down to 
distances very close to black holes and neutron stars.
The LOFT Wide Field Monitor  \cite[WFM,][]{Brandt2012WFM_mn}  will discover and localise X--ray transients and 
impulsive events and monitor spectral state changes with unprecedented sensitivity, 
providing interesting targets for LAD's pointed observations. 
Through the LOFT Burst Alert System (LBAS), the position and occurrence time 
of bright and impulsive events discovered by the WFM will be transmitted on the ground 
within about 30 s from detection.

The LOFT contribution to the SFXT investigation will include unprecedented 
simultaneous high S/N broad-band and time-resolved spectroscopy in several intensity states. 
Starting point for our simulations were \sw\ broad-band observations and detailed {\it XMM-Newton}
observations that describe the outburst state at several intensities 
($F_{\rm (2-10\,keV)}=5.9\times10^{-9}$ to $5.5\times10^{-10}$ erg cm$^{-2}$  s$^{-1}$), 
the intermediate (and most common, $F_{\rm (2-10\,keV)}=10^{-11}$ erg cm$^{-2}$ s$^{-1}$) state, 
and low state ($F_{\rm (2-10\,keV)}=1.2\times10^{-12}$ to $5\times10^{-13}$ erg cm$^{-2}$ s$^{-1}$). 
In our simulations we also considered large variations of $N_{\rm H}$ and the presence of emission lines, 
also as observed by \sw\ and {\it XMM-Newton}. 
For the WFM we used `on axis' ARF, RMF  and the background model of \citet[][]{Gruber1999}, for the 1.5D camera. 
For the LAD we used the `requirements'  (v4) ARF, RMF, and 
background\footnote{User guides at: \url{http://www.isdc.unige.ch/loft/},  under Responses.}.

\begin{figure}
   \includegraphics[angle=270,width=.33\textheight]{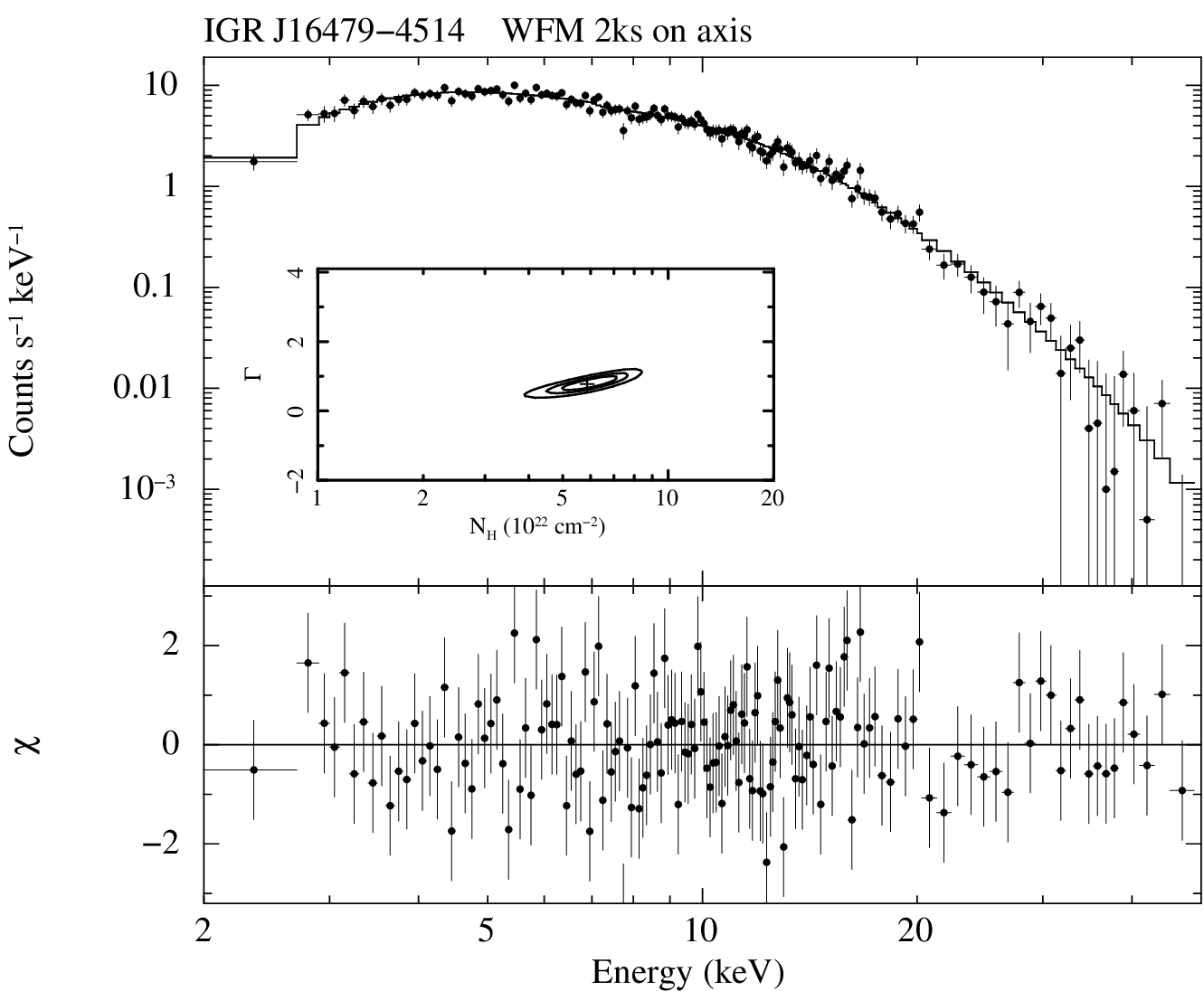}
  \includegraphics[angle=270,width=.33\textheight]{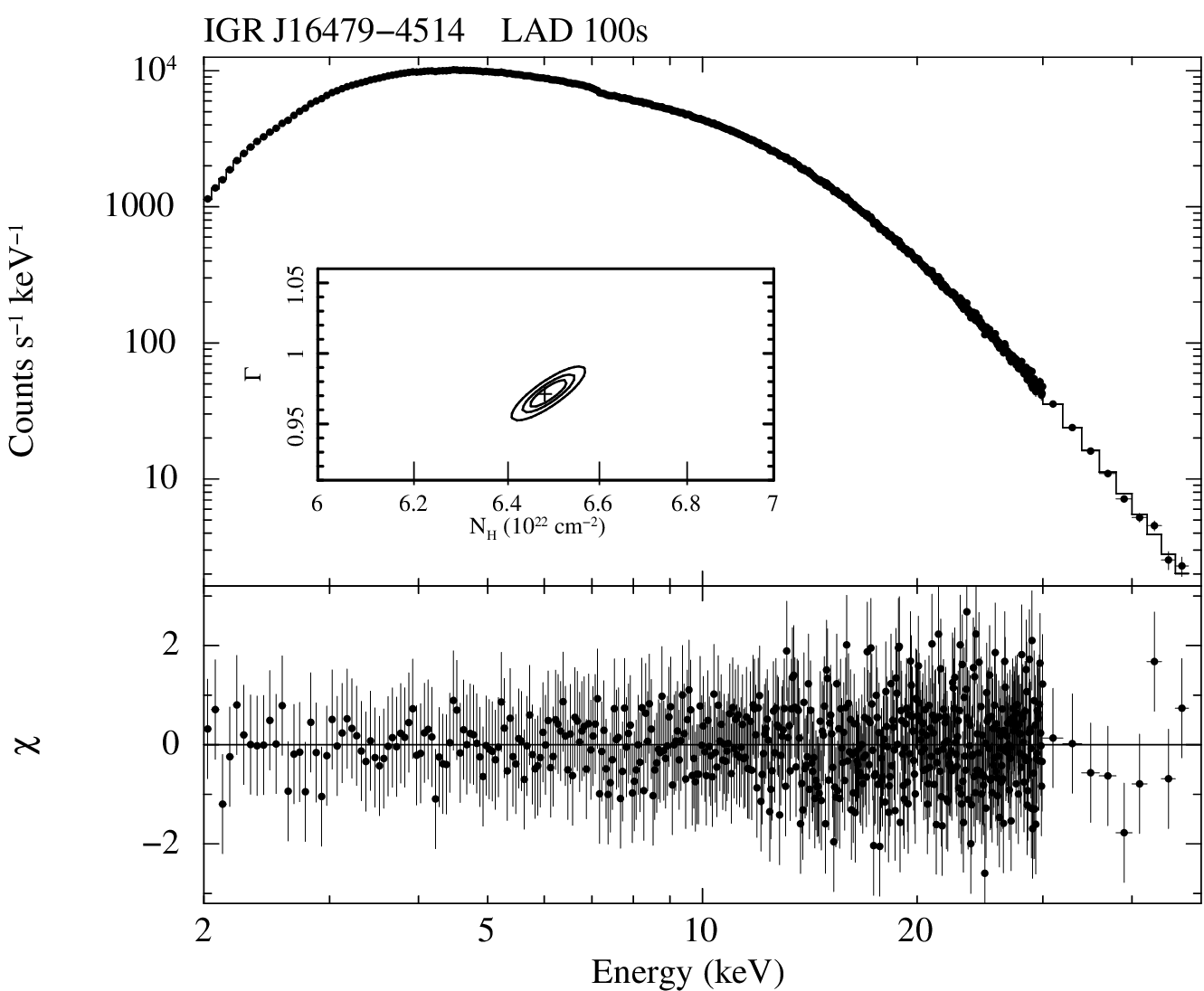}
  \caption{Simulated LOFT spectra of IGR~J16479$-$4514 with an absorbed power-law model with a cut-off  
\citep[][$N_{\rm H} = 6.5 \times10^{22}$ cm$^{-2}$, 
$\Gamma = 0.98$, $E_{\rm c}  =13.5$ keV, $F_{\rm (2-10\,keV)} = 5.9\times10^{-9}$ 
erg cm$^{-2}$ s$^{-1}$]{Romano2008:sfxts_paperII}. 
{\it Left panel:} WFM spectrum (2\,ks).  
 {\it Right panel:} LAD spectrum (100\,s).  
}
  \label{fig12}
\end{figure} 

             \subsection{SFXTs with the WFM}

The sensitive, long term monitoring with the WFM will yield new determinations of 
the SFXT orbital period $P_{\rm orb}$. 
Our simulations show that the WFM will also be able to detect all SFXT short flares within its 
field of view  (FOV) down to a 15--20 mCrab in 5\,ks. 
The calculation of the percentage of SFXT outbursts/flares observed instantaneously in the 
WFM FOV is under investigation. 

The WFM is ideal to catch bright outbursts that reach $6\times10^{-9}$ erg cm$^{-2}$ s$^{-1}$: 
we obtain $\Delta N_{\rm H}/N_{\rm H}$ and $\Delta \Gamma/\Gamma$ within $\sim30$\,\% in 1\,ks, 
within $\sim20$\,\% in 2\,ks (such intense fluxes are maintained for a few ks only, so there is 
no point in investigating longer integrations).
For intermediate flares ($F_{\rm (2-10\,keV)}=10^{-9}$ erg cm$^{-2}$ s$^{-1}$) we obtain $\Delta N_{\rm H}/N_{\rm H}$ 
and $\Delta \Gamma/\Gamma$  $\ga 50$\% in 5ks, while 
flares with fluxes lower than $F_{\rm (2-10\,keV)}=10^{-9}$ erg cm$^{-2}$ s$^{-1}$ 
might require longer exposure times to have sufficiently detailed measurements of the source spectral parameters.
Furthermore, we need $N_{\rm H}\ga 5\times10^{22}$  cm$^{-2}$ to constrain it adequately, given the 2\,keV threshold
with exposure times as low as 2--5\,ks.  
In Fig.~\ref{fig12} (left), we show a 2\,ks simulated WFM spectrum of the SFXT IGR~J16479$-$4514 as 
observed by \sw\ \citep{Romano2008:sfxts_paperII}, 
an absorbed power-law model with a cut-off  ($N_{\rm H} = 6.5 \times10^{22}$ cm$^{-2}$, 
$\Gamma = 0.98$, $E_{\rm c}  =13.5$ keV, $F_{\rm (2-10\,keV)} = 5.9\times10^{-9}$ erg cm$^{-2}$ s$^{-1}$). 

             \subsection{SFXTs with the LAD}

\begin{figure} 
  \includegraphics[angle=270,width=.33\textheight]{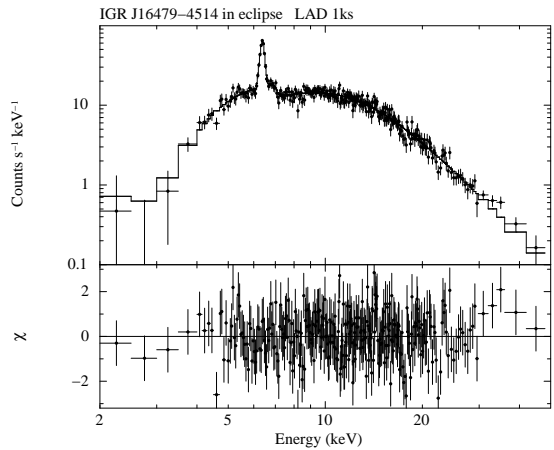}
  \caption{LAD simulated spectrum  (1\,ks)  of  IGR~J16479$-$4514 during eclipse 
\citep[][$N_{\rm H} = 35.2 \times10^{22}$ cm$^{-2}$, 
$\Gamma = 0.98$, 
Norm($K_{\alpha}) =4.6\times10^{-5}$ ph cm$^{-2}$ s$^{-1}$, 
$F_{\rm (2-10\,keV)} = 10^{-11}$ erg cm$^{-2}$ s$^{-1}$]{Bozzo2008:eclipse16479}. 
}
 \label{fig3}
\end{figure}

The LAD is best suited for pointed observations and re-pointing to lower fluxes, 
and line spectroscopy. 
One may be lucky to catch bright outbursts ($F_{\rm (2-10\,keV)} = 6\times10^{-9}$ erg cm$^{-2}$ s$^{-1}$) 
which will yield excellent time-resolved spectroscopy (down to 1s) 
with $\Delta N_{\rm H}/N_{\rm H}$ and $\Delta \Gamma/\Gamma$ within $\sim1$\,\% in 200\,s. 
In Fig.~\ref{fig12} (right), we show a 100\,s LAD simulated spectrum of IGR~J16479$-$4514.  
Intermediate flares will yield  $\Delta N_{\rm H}/N_{\rm H}$ and $\Delta \Gamma/\Gamma$ within 
$\sim5$\,\% in 1\,ks for low $N_{\rm H}$  ($F_{\rm (2-10\,keV)}\sim 9\times10^{-10}$  erg cm$^{-2}$ s$^{-1}$), 
$\Delta N_{\rm H}/N_{\rm H}$ and $\Delta \Gamma/\Gamma$ within $\sim5$\,\% in  1\,ks 
($F_{\rm (2-10\,keV)} = 5\times10^{-10}$ erg cm$^{-2}$ s$^{-1}$).  
Far more likely will be the case of the intermediate state \cite{Romano2011:sfxts_paperVI} 
characterized by fluxes of $F_{\rm (2-10\,keV)} = 10^{-11}$ erg cm$^{-2}$ s$^{-1}$. 
Emission lines can be recovered quite nicely, as shown in Fig.~\ref{fig3}  by the 1\,ks simulated 
spectrum of IGR~J16479$-$4514  in quiescence 
\cite[][note that the exposure time of the {\it XMM--Newton} data where the iron line 
was found was of 28\,ks]{Bozzo2008:eclipse16479}. 
Low intensity states down to $F_{\rm (2-10\,keV)} = 1.2\times10^{-12}$ erg cm$^{-2}$ s$^{-1}$ 
can be studied in 10\,ks.

             \subsection{Pulsed Fractions}
Several SFXTs are X-ray pulsars (with $P_{\rm spin}\sim4$--2000\,s) so it is paramount to determine 
if all of them are indeed pulsating sources and to 
what limit we can determine $P_{\rm spin}$ for the whole sample.
In conjunction with the WFM monitoring, this should in turn allow us to address 
the issue of why SFXTs differ from ordinary HMXBs with the same $P_{\rm orb}$ and $P_{\rm spin}$. 
In particular, we investigated whether we can measure pulsations during the bright outbursts with the WFM, 
and the required length of the pointed LAD observations to accurately measure a $P_{\rm spin}$.
We assumed a sinusoidal pulse profile and Fourier transform searches with $2^{23}$--$2^{24}$ sampled frequencies. 
From known discrete Fourier transform properties, the relationship between the pulsed fraction PF of the 
smallest detectable signal and the number of counts is given by PF$\propto (N_{\rm ph})^{-1/2}$. 
Based on our spectral simulations, 
we expect to be able to detect signals with pulsed fraction as low as those reported in Table~\ref{tab1}. 
Note that red noise, not accounted for in our simulations yet, might lower the sensitivity 
toward the longer periods.

\begin{table} 
\begin{tabular}{lll}
\hline
 & $F_{\rm (2-10\,keV)}$ (erg cm$^{-2}$ s$^{-1}$) & \% \\
 \hline
WFM (5ks) \hspace{0.5truecm} &  $6\times10^{-9}$  \hspace{0.5truecm}   &       2.4 (5$\sigma$) \\
                                 &  $6\times10^{-10}$    &      66  (5$\sigma$), 56 (3$\sigma$) \\

LAD                           &  $6\times10^{-9}$     &  0.08 (5\,ks, 5$\sigma$) \\
                                 &  $6\times10^{-10}$   &  0.3   (5\,ks, 5$\sigma$) \\
                                 &  $1\times10^{-12}$   &  53.4 (10\,ks ,3$\sigma$) \\
\hline
\end{tabular}
\caption{Pulsed fractions as a function of flux.}
\label{tab1}
\end{table} 


\begin{theacknowledgments}
We acknowledge financial contribution from the contract ASI-INAF I/004/11/0.
\end{theacknowledgments}






\IfFileExists{\jobname.bbl}{}
 {\typeout{}
  \typeout{******************************************}
  \typeout{** Please run "bibtex \jobname" to optain}
  \typeout{** the bibliography and then re-run LaTeX}
  \typeout{** twice to fix the references!}
  \typeout{******************************************}
  \typeout{}
 }

\end{document}